\documentclass[aps,pra,reprint,nofootinbib,superscriptaddress,twocolumn,showpacs,showkeys,longbibliography,amsmath,amssymb]{revtex4-2}
\usepackage{graphicx}
\usepackage{dcolumn}
\usepackage{bm}
\usepackage{braket}
\usepackage{subfigure}
\usepackage[colorlinks,bookmarks=false,citecolor=blue,linkcolor=red,urlcolor=blue]{hyperref}
\usepackage[english]{babel}
\usepackage{changes}

\begin{document}
\preprint{APS/123-QED}

\title{Interaction-Induced Dimensional Crossover through Full 3D to 1D}
\author{Tao Yu }
\affiliation{Department of Physics, Zhejiang Normal University, Jinhua 321004, China}
\author{Xiaoran Ye }
\affiliation{Department of Physics, Zhejiang Normal University, Jinhua 321004, China}
\author{Zhaoxin Liang }\email[Corresponding author:~] {zhxliang@zjnu.edu.cn}
\affiliation{Department of Physics, Zhejiang Normal University, Jinhua 321004, China}
\date{\today}

\begin{abstract}

The exploration of dimensional crossover carries profound fundamental significance, serving as a crucial bridge in comprehending the remarkable disparities observed in transitional phenomena across the two distinct dimensions of a physical system.
The prevalent strategy for manipulating the dimensionality involves meticulously controlling the external trapping geometry, thereby 
restricting the degrees of freedom of the kinetic energy from three-dimensional (3D) to lower-dimensional spaces, while maintaining the 3D nature of the interaction energy degrees of freedom. 
The aim of this work is to introduce an innovative scenario to achieve dimensional crossover, characterized by lower-D nature of both the kinetic and the interaction energy degrees of freedom. 
To accomplish this objective, we delve deeply into the realm of a 2D optically trapped Bose gas, focusing specifically on its finite-range interaction. Our emphasis lies in exploring the lattice-induced dimensional crossover from full 3D to 1D in both kinetic and interaction terms.
Utilizing the functional path integral method, we derive the equation of states of the model system, encompassing crucial quantities, such as the ground state energy and quantum depletion. These equations enable us to analyze the combined effects of finite range interaction and an optical lattice on quantum fluctuations of the BEC system. 
Notably, our analytical findings reconcile the Lee-Huang-Yang (LHY) correction to the ground state energy in 3D and Lieb-Liniger (LL) ones in 1D limit, thereby providing fresh insights into the intriguing disparities between LHY and LL corrections.
\end{abstract}
\maketitle

\section{INTRODUCTION}\label{intro}

Currently, there $\textcolor{blue}{\text{are}}$ significant and ongoing interests in exploring how the dimensionality of a physical system impacts its fundamental nature and properties~\cite{Bloch2008}. The motivation behind this exploration is twofold. Firstly, lower dimensions often amplify quantum and thermal fluctuations within the model system, leading to a rich array of genuine quantum effects as dimensionality decreases~\cite{Florian2020,Zhang2020,Ozawa2019}. A typical example consists $\textcolor{blue}{\text{of}}$ the interacting bosons, characterized by the superfluids in three dimension (3D)~\cite{Pitaevskii2016},  Berezinski-Kosterlitz-Thouless (BKT) topological phase transitions in 2D~\cite{Kosterlitz1973,Hadzibabic2006} and  Tomonaga-Luttinger liquids in 1D~\cite{Giamarchi2004} repectively.
Secondly, cutting-edge experimental techniques provide remarkable precision in controlling key parameters of the physical system, such as interaction strength and confining potentials~\cite{Bloch2008}. In particular, the $\textcolor{blue}{\text{system's}}$ dimensionality can be set freely rather than being fixed once for all. In the context of ultracold gases,  highlights in 1D and 2D include the observation of bosonic fermionization into the Tonks-Girardeau (TG) state~\cite{Paredes2004,Meinert2017,Haller2009}, the expansion dynamics~\cite{Roy2024} and the Super-Tonks-Girardeau quench~\cite{Molignini2024} of dipolar bosons. The stark contrast between the aforementioned properties in lower dimensions and their 3D counterparts underscores the importance of exploring the dimensional crossover between these two distinct dimensions.

Along this research line, considerable efforts have been dedicated to gaining a profound understanding of the dimensional crossovers, ranging from a higher to a lower dimension, within the realm of ultracold quantum systems. For instance, Refs.~\cite{Cazalilla2006,Kang2022,Kangkang2022,Zin2018,Ilg2018} have delved into the dimensional crossover exhibited by Bose gas, as well as the counterpart in Fermi gas~\cite{Dyke2011,Peppler2018,Shi2023,Gong2023}. Through the meticulous manipulation of the transverse $z$-direction, Refs.~\cite{Hu2011,Faigle-Cedzich2021,Sommer2012,Keepfer2022,Pekker2010} have successfully achieved a dimensional crossover from 3D to 2D. Furthermore, Refs.~\cite{Orso2006,Ying2009} have demonstrated that the introduction of an external trapping potential gives rise to a crossover from 3D to 1D. It is noteworthy that recent theoretical and experimental~\cite{Yanliang2024,Guo2023,Hepeng2023} investigations have revealed the dimensional crossover of strongly-interacting bosons from 2D to 1D, highlighting the rich and complex nature of these systems across different dimensions. 

Dimensional crossovers are distinguished by their hierarchical access to novel energy and length scales. Leveraging this understanding, the predominant approach $\textcolor{blue}{\text{to}}$ realizing dimensional crossover during the last decade is that a physical system becomes quasi-low dimensional when energetic restriction to freeze excitations in one or two dimensions is reached, which is referred $\textcolor{blue}{\text{to as}}$ the kinetic-induced low-dimensional physical system~\cite{Orso2006,Ying2009,Yanliang2024,Hepeng2023,Guo2023,Umesh2024}. It is important to note that the interactions in such low-dimensional systems are often treated analogously to their 3D counterparts. However, in the following, we introduce an alternative scenario based on a 2D optically-trapped Bose gas with weak interactions and specifically engineered finite-range interatomic interactions. As we will demonstrate, considering finite-range interactions offers a novel perspective on low-dimensional systems, characterized by both their interactions and kinetic energy being inherently low-dimensional.

The second impetus for this work stems from recent experimental advancements in realizing ultracold quantum gases with finite-range interatomic interaction~\cite{Hai2012,Haibin2012}. $\textcolor{blue}{\text{Compared}}$ with the quantum gases exhibiting s-wave interactions, ultracold quantum gases with finite-range interatomic interactions possess crucial novel features, as outlined below: The equation of state (EOS) for s-wave interacting quantum gas exhibits a profound universality, stemming from the fact that it can be precisely characterized by a single parameter: the s-wave scattering length $a_{\text s}$. This parameter serves as a versatile tool, encapsulating not just the fundamental nature of two-body interactions but also the intricate many-body physics~\cite{Lee1957,LeeTD1957} of the system. However, the introduction of finite-range atomic interactions disrupts this universality, giving rise to nonuniversal effects within the EOS~\cite{Braaten2001,Fu2003,ACappellaro2017,Lorenzi2023,Ralf2006}.  At the mean-field level, the Gross-Pitaevskii equation undergoes modifications~\cite{Collin2007,Sgarlata2015,Fu2003,Veksler2014} to incorporate these nonuniversal effects. The modified Gross-Pitaevskii equation provides a comprehensive framework for elucidating the physical behavior of nonuniform condensates, effectively capturing deviations from the idealized s-wave interaction scenario. Beyond the mean-field approximation, recent studies have broadened the thermodynamic analysis of pure two-dimensional and three-dimensional uniform Bose gases~\cite{Cappellaro2017,Salasnich2017} to incorporate finite-range interactions, reaching the sophisticated Gaussian level of approximation. We mention that our previous work~\cite{Xiaoran2024} has investigated finite-range-interaction induced EOS along a dimensional crossover from 3D to 2D. These works provide a deeper comprehension of  the rich and intricate physics that arises in the presence of non-ideal interactions.

In this work, we have delved into the intricate phenomenon of dimensional crossover exhibited by Bose gases with finite-range interactions confined within optical lattices. Employing the finite-temperature functional path integral framework and the tight-binding approximation, we have embarked on a comprehensive exploration of the ground-state properties of these Bose gases in a two-dimensional optical lattice setting. Our findings have culminated in the derivation of a beyond-mean-field equation of state for the Bose system, one that encompasses the influence of various parameters, including optical lattice configurations, effective ranges, and scattering lengths. This equation of state yields generalized corrections to the Lee-Huang-Yang (LHY) energy in 3D and the Lieb-Liniger (LL) solution in 1D. Furthermore, by manipulating the effective range, we have visually analyzed the subsequent influence of finite-range interactions on Gaussian quantum fluctuations, offering a deeper insight into their pivotal role in dimensional crossover phenomena.

The structure of this paper is outlined as follows: In Sec.~\ref{sec2}, we introduce the action functional of the model system and demonstrate the fundamental terminology of the finite-temperature functional path integral. In Sec.~\ref{sec3}, we calculate the ground state energy and quantum depletion of the system through thermodynamic relationships and the regularization method. Additionally, we analyze the asymptotic behavior of these two physical quantities under both the 1D and 3D limits. Sec.~\ref{conclu} summarizes our work, and we also consider the potential experimental conditions for realizing our proposed scenario.

\section{Interaction-Induced Dimensional Crossover}\label{sec2}

A common strategy $\textcolor{blue}{\text{for}}$ achieving the low-D system in an ultra-cold quantum gas is to add an optical lattice to the microscopic Hamiltonian~\cite{Orso2006,Hu2011}. To illustrate this, let's consider the dimensional crossover from 3D to quasi-1D by introducing a 2D optical lattice as an example. The dispersion of single-particle energy in the model system is given by $\epsilon_{\mathbf{k}}^{0}=\hbar^{2}k_{z}^{2}/2m+J\left[2-\cos k_{x}d-\cos k_{y}d\right]$ with $J$ being the tunneling rate between two optical wells. When the energetic constraints freeze excitations along $x$- and $y$- directions, the system effectively becomes quasi-1D with the energy dispersion simplified into $\epsilon_{\mathbf{k}}^{0}=\hbar^{2}k_{z}^{2}/2m$ as $J\rightarrow 0$. Transitioning into the quasi-1D regime, the coupling constant is renormalized as $\tilde{g}=C^2g_0$ with $g_0$ being the 3D coupling constant and $C$ being the renormalized constant~\cite{Orso2006}.  We refer $\textcolor{blue}{\text{to it as}}$ the kinetic-induced 1D physical system, distinguished by its 1D kinetic energy and 3D interaction energy.

Below, we embark on a strategy to engineer the form of inter-atomic interaction, aiming to achieve a dimensional crossover from 3D to quasi-1D for both the kinetic and interaction energy of the model system. This crossover, dubbed the interaction-induced 1D system, involves meticulously selecting the inter-atomic  pseudo-potential~\cite{Roth2001,Fu2003}  to facilitate this transition
\begin{equation}
V\left(r\right)=g_{0}\delta\left(r\right)-\frac{g_{2}}{2}\left[\overleftarrow{\nabla}^{2}\delta\left(r\right)+\delta\left(r\right)\overrightarrow{\nabla}^{2}\right],\label{PsedoP}
\end{equation}
with $g_{0}=4\pi\hbar^{2}a_{s}/m$ and $g_{2}=2\pi\hbar^{2}a_{s}^{2}r_{s}/m$ being the s-wave scattering and finite-range coupling $\textcolor{blue}{\text{constants}}$. Here, $a_{s}$ is the s-wave scattering length and $r_{s}$ $\textcolor{blue}{\text{represents}}$ the effective range of interatomic interaction potential~\cite{Dalfovo1999}.  With the introduction of a 2D optical lattice, 
the interaction energy corresponding to Eq.~(\ref{PsedoP}) is proportional to be $\tilde{g}_2\left[k_{z}^{2}+J_{1}\left(2-\cos k_{x}d-\cos k_{y}d\right)\right]$. By changing $J_1\rightarrow 0$, one can reach the regime of 1D interaction $\textcolor{blue}{\text{in}}$ the form of  $\tilde{g}_2k_{z}^{2}$.  In what follows, we focus on the equation of state of the interacting Bose gas trapped $\textcolor{blue}{\text{in}}$ a 2D optical lattice and demonstrate the interaction-induced dimensional crossover through full 3D to qusi-1D.

We adopt the path-integral approach in the study of the weakly interacting Bose gas with the finite range interaction in a 2D optical lattice. The grand canonical partition function $\mathcal{Z}$ of the system has the form~\cite{Nagaosa1999,Atland2010}
\begin{equation}
\mathcal{Z}=\int D\left[\psi,\psi^{*}\right]\exp\left[-\frac{S\left[\psi,\psi^{*}\right]}{\hbar}\right],\label{PartFun}
\end{equation}
where
\begin{widetext}
\begin{equation}
S\left[\psi,\psi^{*}\right]=\int_{0}^{\hbar\beta}d\tau\int
d\mathbf{r}\psi^{*}\left(\mathbf{r},\tau\right)\left[\hbar\frac{\partial}{\partial\tau}-\frac{\hbar^{2}\nabla^{2}}{2m}-\mu+V_{\text{\text{opt}}}\left(\mathbf{r}\right)\right]\psi\left(\mathbf{r},\tau\right)+\frac{g_{0}}{2}\left|\psi\right|^{4}-\frac{g_{2}}{2}\left|\psi\left(\mathbf{r},\tau\right)\right|^{2}\nabla^{2}\left|\psi\left(\mathbf{r},\tau\right)\right|^{2},\label{ActFun}
\end{equation}
\end{widetext}
is the action functional and bosonic atoms described by the complex field $\psi\left(\mathbf{r},\tau\right)$, and $\tau$ is imaginary time, $\beta=1/k_{B}T$ with $k_{B}$ being the Boltzmann constant and $T$ being the temperature.

The 2D optical lattice of $V_{\text {opt}}({\bf r})$ in action functional (\ref{ActFun}) reads~\cite{Bloch2005}
\begin{equation}
V_{\text{opt}}\left(\mathbf{r}\right)=t\times E_{R}\left[\sin^{2}\left(q_{B}x\right)+\sin^{2}\left(q_{B}y\right)\right],\label{OptPoten}
\end{equation}
where $t$ describes the dimensionless intensity of the laser beam and $E_{R}$ is the recoil energy with $q_{B}$ being the Bragg momentum~\cite{Bloch2008}. The lattice period is fixed by $d=\pi/q_{B}$ with $d$ being the lattice spacing. Atoms are unconfined in the $z$ plane. By incorporating a 2D optical lattice, as described in Eq. (\ref{OptPoten}), into a BEC existing in a uniform spatial configuration, which triggers a dimensional crossover from 3D to quasi-1D~\cite{Orso2006,Hu2011}, we enable graded access to novel energy and length scales. Note that the equation of state for BEC exhibiting finite-range effective interaction, which aligns with Eq. (\ref{ActFun}) while $V_{\text {opt}}({\bf r})=0$, has been meticulously derived and comprehensively presented in Refs.~\cite{Cappellaro2017,Salasnich2017}. Meanwhile, Refs. \cite{Orso2006,Hu2011} have investigated the lattice-induced dimensional crossover from 3D to quasi-1D for a BEC without finite-range effective interaction, corresponding to Eq. (\ref{ActFun}) with $g_2=0$

Additionally, in contrast to three-dimensional free Bose gases, the effective coupling constant in Eq.~(\ref{ActFun}) is intricately linked to the tight constraints imposed by the optical lattice in a particular direction. In the presence of an optical lattice, both the s-wave coupling constant and the finite-range coupling constant are jointly determined by the lattice constant and the density. It is worth emphasizing that in our work, we abstained from considering the impact of confinement-induced resonance (CIR)~\cite{Peng2010,Zhang2011} on the coupling constant. Feshbach resonance~\cite{Bergeman2003} offers a profound framework for explaining the underlying physics of CIR, where the ground-state transverse mode and the other transverse modes along the tightly constrained dimension correspond to the scattering open channel and closed channel, respectively. The tight-binding approximation employed in our study confines the ultracold atoms to the lowest Bloch band, thereby eliminating the existence of transverse modes along the tight-confinement dimension. Consequently, the influence of CIR can be safely disregarded.

\textcolor{blue}{We proceed to provide approximate estimates of the parameter ranges where the tight-binding approximation remains valid. For this purpose, we adopt the commonly encountered experimental parameters for an optically trapped Bose gas reported in Ref.~\cite{Du2010}. It is crucial to note that the tight-binding approximation holds under specific conditions. Primarily, the lattice depth $t$ in Eq. (\ref{OptPoten}) must be relatively large $\left(t \geq 5\right)$ to ensure that the interband gap $E_{\text{gap}}$ exceeds the chemical potential $\mu$~\cite{Orso2006,Hu2011,Ying2009,Zhou2010}. Secondly, under such conditions, numerous isolated wells are formed, giving rise to an orderly array of condensates. Simultaneously, due to quantum tunneling, the overlap between the wave functions of two continuous wells is still sufficient to ensure full coherence. Additionally, the typical detailed parameters read as follows: The recoil energy is $E_{R} \approx h \times 3.33$ kHz with $h$ being the Plank constant and the chemical potential of gas is $\mu \approx \tilde{g}_{0}n_{0} \approx h \times 400$ kHz. In the case of $t \approx 5$, we can estimate the parameters of $J / E_{R} \approx 0.18$ based on Ref.~\cite{Likharev1985}. Consequently, the dimensionless parameters used in the figures of this work are as follows: $s=2 J/\tilde{g}_{0}n_{0} \approx 3.6$, indicating that our model system exhibits 3D-like. Furthermore, as shown in ~\cite{Du2010}, the optically-trapped Bose gas exhibits complete superfluidity below the critical lattice height $t_{c} \approx 13$ corresponding to $J$ and $J_{1}$ being nearly zero~\cite{Likharev1985}. We, therefore, conclude that the tight-binding approximation can be regarded as valid within the range of $5<t<13$, corresponding to $0<s<4$ as depicted in Fig. \ref{figure1}. Based on this assumption, we refer to the density of the condensate as $n_0$ and ignore the phase transition of the Mott insulator.}

Following Ref.~\cite{Hu2011,Ying2009,Zhou2010}, we treat our model system within the tight-binding approximation. Within this framework, the lowest Bloch band of the system can be accurately described in terms of the Wannier functions as $\phi_{k_{x}}\left(x\right)\phi_{k_{y}}\left(y\right)$ with $\phi_{k_{x_{i}}}\left(x_{i}\right)=\sum_{l}e^{ildk_{x_{i}}}w\left(x_{i}-ld\right)$ and $w\left(x_{i}\right)=\text{\ensuremath{\sqrt{d}}\ensuremath{\exp\left[-x_{i}^{2}/2\sigma^{2}\right]}}/\pi^{1/4}\sigma^{1/2}$ with $d/\sigma=\pi t^{1/4}\exp\left(-1/4\sqrt{t}\right)$ $\left(i=1,2\text{ and }x_{1}=x,x_{2}=y\right)$~\cite{Orso2006}. Expanding the bosonic field variables in (\ref{ActFun}) by the expression
$\psi\left(\mathbf{r},\tau\right)=\sum_{\mathbf{k},n}\psi_{\mathbf{k},n}\phi_{k_{x}}\left(x\right)\phi_{k_{y}}\left(y\right)e^{-ik_{z}z}e^{i\omega_{n}\tau}$ with the bosonic Matsubara frequencies $\omega_{n}=2\pi n/\hbar\beta$, we obtain
\begin{eqnarray}
&&\frac{S\left[\psi,\psi^{*}\right]}{\hbar\beta V} =\sum_{\mathbf{k},n}\psi_{\mathbf{k},n}^{*}\left[-i\hbar\omega_{n}+\epsilon_{\mathbf{k}}^{0}-\mu\right]\psi_{\mathbf{k},n}\nonumber \\
 && +\frac{\tilde{g}_{0}}{2}\sum_{{\mathbf{k},\mathbf{k^{\prime}},\mathbf{q}\atop n,n^{\prime},m}}\psi_{\mathbf{k}+\mathbf{q},n+m}^{*}\psi_{\mathbf{k^{\prime}-\mathbf{q}},n^{\prime}-m}^{*}\psi_{\mathbf{k^{\prime}},n^{\prime}}\psi_{\mathbf{k},n}\nonumber \\
 && +\frac{\tilde{g}_{2}}{2}\sum_{{\mathbf{k},\mathbf{k^{\prime}},\mathbf{q}\atop n,n^{\prime},m}}\tilde{q}^{2}\psi_{\mathbf{k}+\mathbf{q},n+m}^{*}\psi_{\mathbf{k^{\prime}-\mathbf{q}},n^{\prime}-m}^{*}\psi_{\mathbf{k^{\prime}},n^{\prime}}\psi_{\mathbf{k},n},\label{FourierActfun}
\end{eqnarray}
where $\epsilon_{\mathbf{k}}^{0}=J(2-\cos k_{x}d-\cos k_yd)+\hbar^2k_z^2/2m$ is the energy dispersion of the non-interacting system and $\tilde{q}^{2}=k_{z}^{2}+J_{1}\left(2-\cos k_{x}d-\cos k_{y}d\right)$ is the correction of momentum exchange between particles with $J=-2/d\int_{0}^{d}dxw^{*}\left(x\right)\left(-\hbar^{2}\partial_{x}^{2}/2m+V_{\text{\text{opt}}}\right)w\left(x-d\right)$  and $J_{1}=2\sqrt{2\pi}\sigma/d^{2}\int_{0}^{d} dxw^{2}\left(x-d\right)\partial_{x}^{2}w^{2}\left(x\right)$.
Based on the functional of Eq. (\ref{FourierActfun}),  we can demonstrate the interaction-induced dimensional crossover proposed in this work explicitly as follows:

Firstly,   the functional of Eq. (\ref{FourierActfun}) has two importances with respect to the functional characterizing a trapped Bose gas in the absence of optical confinement: The kinetic energy in the first line of the functional of Eq. (\ref{FourierActfun}) exhibits
a periodic dependence as $\varepsilon^0_{\mathbf{k}}=J(2-\cos k_{x}d-\cos k_yd)+\hbar^2k_z^2/2m$. In the 3D limit when the lattice strength is weak, the kinetic energy can be rewritten as $\varepsilon^0_{\mathbf{k}}=\hbar^2 (k_x^2+k_y^2)/2m^*+\hbar^2k_z^2/2m$ with $m^*=\hbar^2/Jd^2$.  With the increase of lattice strength, the value of $J$ will decay exponentially. As a result, the kinetic energy experiences a dimensional crossover from 3D to 1D and becomes to be an effective 1D one of $\varepsilon^0_{\mathbf{k}}=\hbar^2k_z^2/2m$. (ii) The s-wave scattering coupling constant in the second line of the functional of Eq. (\ref{FourierActfun}) is renormalized as $\tilde{g}_0=C^2g_0$ with $C=\sqrt{\pi/2}t^{1/4}e^{-1/4\sqrt{t}}$.

Secondly, the finite-range interaction term in the third line of  the functional of Eq. (\ref{FourierActfun}) exhibits the dynamical 1D features of particle motion, and a dimensional crossover from 3D to 1D emerges in the behavior of the interaction energy when the energetic restriction to freeze anxial excitations is reached.

(i) For $8J/\mu \gg 1$, the system exhibits a distinct anisotropic 3D behavior, and the finite-range effective coupling constant takes $\textcolor{blue}{\text{the}}$ form of $\frac{\tilde{g}_2\tilde{q}^2}{2}=\frac{g_2C^2m}{\hbar^2}\times \left[\frac {\hbar^2(k_x^2+k^2_y)}{2m(2/J_1d^2)}+\frac {\hbar^2k_z^2}{2m}\right]$. 

(ii) For $8J/\mu\ll 1$,   the finite-range interaction term in the third line of  the functional of Eq. (\ref{FourierActfun}) exhibits a dimensional crossover from 3D to 1D with the form of $\frac{\tilde{g}_2\tilde{q}^2}{2}=\frac{g_2C^2m}{\hbar^2}\times \left[\frac {\hbar^2k_z^2}{2m}\right]$, corresponding to $J_1\rightarrow 0$.

Now, we are ready to calculate the analytical expressions of the equation of state of $\textcolor{blue}{\text{the}}$ model system along the dimensional crossover. In this end, by applying the mean field plus Gaussian (one loop) approximation~\cite{Cappellaro2017} to Eq.~(\ref{FourierActfun}) and writing $\sum_{\mathbf{k},n}\psi_{\mathbf{k},n}=\sqrt{n_{0}}+\sum_{\mathbf{k},n\neq0}\phi_{\mathbf{k},n}$, we find that the Gaussian contribution of quantum fluctuation is described by 
\begin{equation}
S_{g}\left[\phi,\phi^{*}\right]=\frac{1}{2}\sum_{\mathbf{k},n}\left(\begin{array}{cc}
\phi_{\mathbf{k},n}^{*} & \phi_{\mathbf{-k},-n}\end{array}\right)\mathbf{M}\left(\mathbf{k},n\right)\left(\begin{array}{c}
\phi_{\mathbf{k},n}\\
\phi_{\mathbf{-k},-n}^{*}
\end{array}\right).\label{GaussContri}
\end{equation}
The inverse fluctuation propagator $\mathbf{M}\left(\mathbf{k},n\right)$ is given by the following:
\begin{widetext}
\begin{equation}
\mathbf{M}\left(\mathbf{k},n\right)=\beta\left(\begin{array}{cc}
-i\hbar\omega_{n}+\epsilon_{\mathbf{k}}^{0}-\mu+\tilde{g}_{0}n_{0}+n_{0}\tilde{V}\left(\mathbf{k}\right) & n_{0}\tilde{V}\left(\mathbf{k}\right)\\
n_{0}\tilde{V}\left(\mathbf{k}\right) & i\hbar\omega_{n}+\epsilon_{\mathbf{k}}^{0}-\mu+\tilde{g}_{0}n_{0}+n_{0}\tilde{V}\left(\mathbf{k}\right)
\end{array}\right),\label{Matrix}
\end{equation}
\end{widetext}
where the chemical potential being $\mu=\tilde{g}_{0}n_{0}$ and $\tilde{V}\left(\mathbf{k}\right)=\tilde{g}_{0}+\tilde{g}_{2}\left[k_{z}^{2}+J_{1}\left(2-\cos k_{x}d-\cos k_{y}d\right)\right]$. By integrating on the Boson fields $\phi_{\mathbf{k},n}$ and $\phi_{\mathbf{k},n}^{*}$, the sum over bosonic Matsubara frequencies gives~\cite{Andersen2004} the Gaussian grand potential
\begin{equation}
\Omega_{g}=\textcolor{blue}{\sum_{\mathbf{k}}}\frac{E_{\mathbf{k}}}{2}+\frac{1}{\beta}\ln\left(1-e^{-\beta E_{\mathbf{k}}}\right),\label{GrandPoten}
\end{equation}
where $E_{\mathbf{k}}$ is the energy dispersion:
\begin{equation}
E_{\mathbf{k}}=\sqrt{\epsilon_{\mathbf{k}}^{0}\Big[\epsilon_{\mathbf{k}}^{0}+2n_{0}\tilde{V}\left(\mathbf{k}\right)\Big]}.\label{EnergyDisper}
\end{equation}

Before moving on to the subsequent calculation, we verify the formula (\ref{EnergyDisper}) to ascertain if the energy dispersion can be restored to the existing work when either the optical lattice $V_{\text{opt}}$ disappears or the finite-range interaction $g_{2}$ vanishes. In the limit of $V_{\text{opt}}=0$ and $g_{2}\neq0$, the energy dispersion returns to the corresponding one in Ref.~\cite{Cappellaro2017}, and then in the limit of $V_{\text{opt}}\neq0$ and $g_{2}=0$, Eq.~(\ref{EnergyDisper}) recovers the corresponding one in Ref.~\cite{Ying2009}. When both the optical lattice and finite-range interaction vanish, our findings should be identical with the well-known Bogoliubov spectrum of collective excitations.

\section{Equation of state of model system from full 3D to quasi-1D}\label{sec3}

In the previous Sec. \ref{sec2}, we present the novel protocol $\textcolor{blue}{\text{for}}$ realizing interaction-induced dimensional crossover. In Sec. \ref{sec3}, we plan to obtain the analytical expressions of $\textcolor{blue}{\text{the}}$ equation of state of $\textcolor{blue}{\text{the}}$ model system along the dimensional crossover from 3D to quasi-1D. 

In this work, we are interested in $\textcolor{blue}{\text{the}}$ equation of state at zero temperature. As such, the analytical expression of the ground state energy $E_{g}=\Omega^{0}\left(\mu,n_{0}\right)+\mu N_{0}$ can be obtained as 
\begin{equation}
\frac{E_{g}}{V}=\frac{1}{2}\tilde{g}_{0}n_{0}^{2}+\frac{1}{2V}\sum_{\mathbf{k}}\left[\sqrt{\epsilon_{\mathbf{k}}^{0}\left(\epsilon_{\mathbf{k}}^{0}+2\tilde{V}\left(\mathbf{k}\right)n_{0}\right)}\right].\label{GroundEnergy1}
\end{equation}
Equation (\ref{GroundEnergy1}) contains an ultraviolet divergence in the limit of large momentum,  which can be regularized by the mean of momentum-cutoff regularization (MCR)~\cite{Salasnich2016}. After changing the sum into integrals in Eq. (\ref{GroundEnergy1}) and performing the integration over the axial momentum $k_{z}$, we can derive the analytical expression of ground state energy as 
\begin{equation}
\frac{E_{g}}{V}=\frac{1}{2}\tilde{g}_{0}n_{0}^{2}-\frac{\tilde{g}_{0}n_{0}\sqrt{2m\tilde{g}_{0}n_{0}}}{4\pi\hbar d^{2}}f\left(\frac{2J}{\tilde{g}_{0}n_{0}}\right),\label{GroundEnergy2}
\end{equation}
where the first term represents the mean-field contribution, whereas the second term corresponds to the correction beyond the mean-field approximation, which arises due to quantum fluctuations and the scaling function $f\left(s\right)$ with the variable $s=\ensuremath{2J/\tilde{g}_{0}n_{0}}$
which can be controlled by the strength of $\textcolor{blue}{\text{the}}$ optical lattice, is given by
\begin{equation}
f\left(s\right)=\frac{\pi}{2\left(2\pi\right)^{2}}\int_{-\pi}^{\pi}\!\mathrm{d}^{2}k\Big\{{}_{2}F_{1}\left[\frac{1}{2},\frac{3}{2},3,\frac{-2\nu}{s\gamma}\right]\!\sqrt{\frac{1+\chi\mu}{s\gamma}}\nu^{2}\Big\},\label{ScalingFunF}
\end{equation}
where the integration over the transverse quasimomenta is restricted to the first Brillouin zone $\left|k_{x}\right|,\left|k_{y}\right|\leq\pi$~\cite{Orso2006} and the function $_{2}F_{1}\left[a,b,c\text{,}d\right]$ is the hypergeometric function with $\textcolor{blue}{\text{multiple}}$ parameters, including the finite range factor $\chi=4m\tilde{g}_{2}/\hbar^{2}\tilde{g}_{0}$, the dimensionless second-order tunneling rate $\beta=\hbar^{2}J_{1}/2m\tilde{g}_{0}n$ and several combination parameters $\lambda=1+\chi\mu$, $\gamma=1-\cos{k_{x}}/2-\cos{k_{y}}/2$ and $\nu=\left(2+s\gamma+2\chi\mu\cdot\beta\gamma\right)/2\lambda-s\gamma/2$

Quantum depletion refers to the number of atoms with nonzero momentum which can be calculated through a thermodynamic equation \textcolor{blue}{$n=-\partial_{\mu}\Omega^{\left(0\right)}\left(\mu,n_{0}\right)/V$ with $\Omega^{\left(0\right)}\left(\mu,n_{0}\right)=\Omega_{0}\left(\mu,n_{0}\right)+\Omega_{g}^{\left(0\right)}\left(\mu,n_{0}\right)$. The first term $\Omega_{0}\left(\mu,n_{0}\right)=\left(\frac{1}{2}\tilde{g}_{0}n_{0}^{2}-\mu n_{0}\right)V$ is the mean-field contribution and the second term $\Omega_{g}^{\left(0\right)}\left(\mu,n_{0}\right)=\frac{1}{2}\sum_{\mathbf{k}} E_{\mathbf{k}}\left(\mu,n_{0}\right)$ is the zero temperature contribution of quantum Gaussian fluctuations. From this, we obtain the particle density $n=n_{0}-\frac{1}{2V}\sum_{\mathbf{k}}\partial_{\mu}E_{\mathbf{k}}\left(\mu,n_{0}\right)$. After changing the sum into integrals and using the MCR method again,} we obtain the analytical expression of quantum depletion $\Delta N$ as follows 
\begin{equation}
\frac{\Delta N}{V}=\frac{\sqrt{2m\mu}}{4\pi\hbar d^{2}}h\left(s\right),\label{Depletion}
\end{equation}
with
\begin{eqnarray}
h\left(s\right) &=&\frac{1}{\left(2\pi\right)^{2}}\int_{-\pi}^{\pi}\mathrm{d}^{2}k\int_{0}^{\infty}\sqrt{\frac{\lambda}{z}}\mathrm{d}z\Big[\frac{\left(z+s\gamma\right)\left(1+\frac{1}{\lambda}\right)+\nu}{2\sqrt{\left(z+s\gamma\right)\left(z+s\gamma+\nu\right)}}\nonumber \\
 &-&\frac{1}{2}\left(1+\frac{1}{\lambda}\right)+\frac{\nu}{4\left(z+s\gamma\right)}\left(\frac{1}{\lambda}-1\right)\Big].\label{ScalingFunH}
\end{eqnarray}

\begin{figure}
\begin{centering}
\includegraphics[scale=0.5]{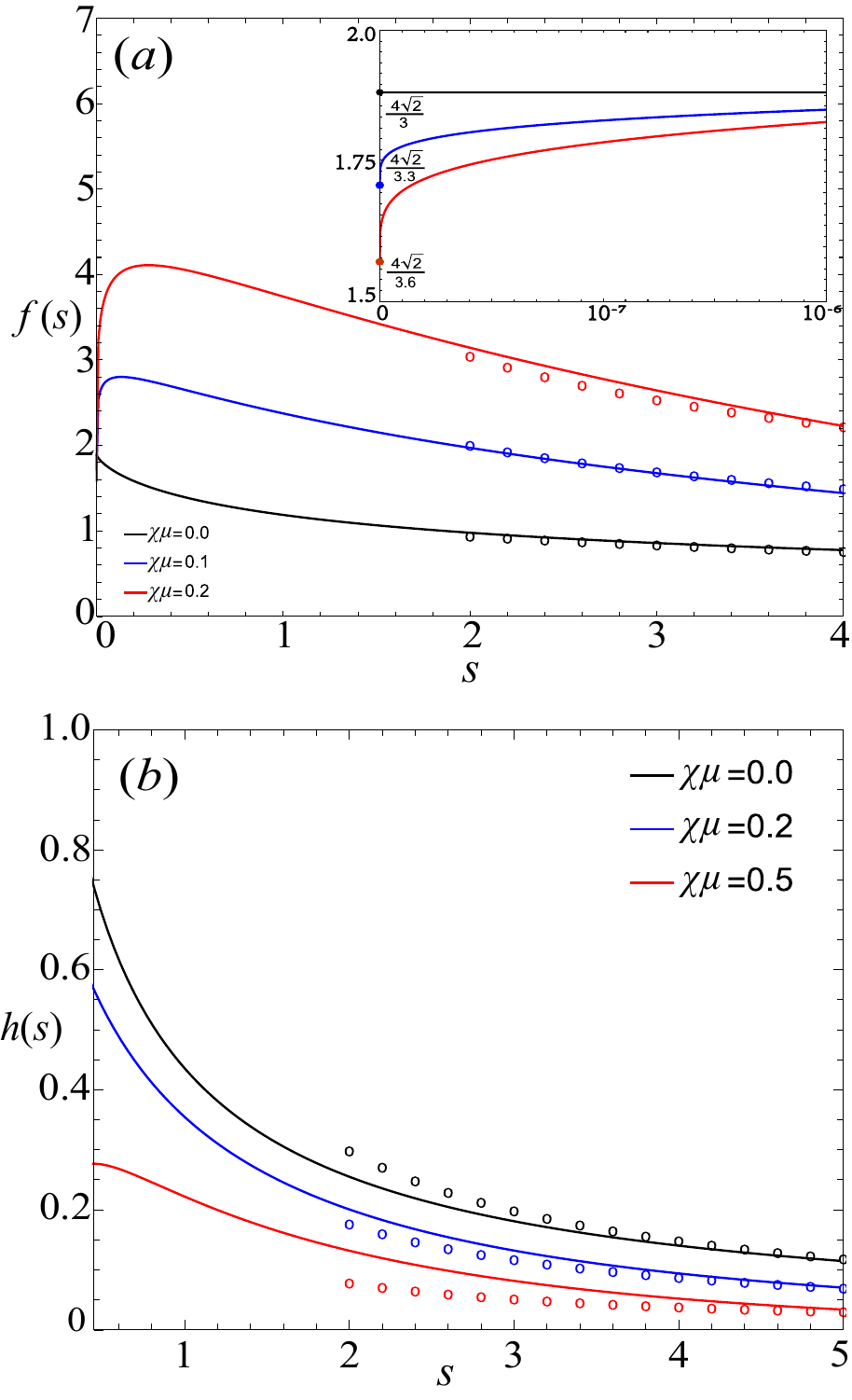}
\par\end{centering}
\caption{$\left(a\right)$ Scaling function $f\left(s\right)$ of Eq. (\ref{ScalingFunF}) (solid line) as function of dimensionless
tunneling rates $s=2J/\tilde{g}_0n_0$ with different values of $\chi\mu$. The dotted curves represent the 3D asymptotic behavior of $f\left(s\right)$ in the regime of $s\gg 1$.
The concrete values of $f\left(s=0\right)=4\sqrt{2}/3\left(1+\chi\mu\right)$ in the 1D limit are displayed by the three dots in the inset. $\left(b\right)$ Scaling function $h\left(s\right)$ of Eq. (\ref{ScalingFunH}) 
(solid line) and its 3D (dotted line) asymptotic behavior as as function of $s$ with different values of $\chi\mu$.}\label{figure1}
\end{figure}

\textcolor{blue}{Prior to delving into the intricacies of ground state energy and quantum depletion, it is imperative to examine the rationality of the value of the dimensionless finite-range coupling constant $\chi\mu=4d^{2}/\sigma^{2}\cdot \left(a^{3}_{s} n_{0}\right)\cdot r_{s}/a_{s}$. This step is fundamental as it informs us about the potential experimental viability of our proposed model. Considering the typical experiments in an optically trapped BEC as in~\cite{Bloch2008}, the relevant parameter $d/\sigma \sim 1$. To illustrate, let us consider the example of ${}^{6} \text{Li}$, as referenced in~\cite{Bartenstein2005}. In this instance, the typical density $n_{0}$ is approximately $4\times 10^{12}\text{cm}^{-3}$.
	Concurrently, the scattering length $a_s$ is estimated to be around $1.13\times 10^{-7}\text{m}$. The order of magnitude of $a^{3}_{s}n_{0}$ is approximately $10^{-3}$. Furthermore, as documented in~\cite{Wu2012}, the effective distance $r_s$ is estimated to be within the range of $0\sim3.71\times10^{-6}\text{m}$. By inserting these parameters into $\chi \mu$, we can approximate its value to be within the range of $0\sim 0.75$. The maximum value of $\chi \mu $ in Fig. \ref{figure1} is within this range.}

Equations  (\ref{GroundEnergy2}) and (\ref{Depletion}) are the key results of our work. Based on Eqs.  (\ref{GroundEnergy2}) and (\ref{Depletion}), we investigate the non-universal equations of state of model system along the dimensional crossover from full 3D to quasi-1D with the emphasis on the finite-range interaction effects $\textcolor{blue}{\text{labeled}}$ by $g_2$. In this end, we plot  the functions of $f\left(s\right)$ and $h\left(s\right)$ of Eqs. (\ref{ScalingFunF}) and (\ref{ScalingFunH}) into Figs. \ref{figure1}(a) and (b) respectively, based on which we can analyze interaction-induced dimensional crossover through full 3D to 1D.

In the 3D limit of $s\gg 1$ corresponding to weak lattice depth limit,  the scaling function (\ref{ScalingFunF}) approaches the asymptotic law $f\left(s\right)\simeq\frac{1.43+9.8\chi\mu}{\sqrt{s/2}}-\frac{32\sqrt{2}}{15\pi{s}}\cdot\frac{1}{\left(1+\chi\mu\right)^{2}}$ as shown by the dotted curves in Fig. \ref{figure1}(a). By plugging such asymptotic behavior of $f\left(s\right)$ into Eq. (\ref{GroundEnergy2}), we can obtain the asymptotic form of the ground state energy in the limit of 3D
\begin{eqnarray}
\frac{E_{g}}{V} =\frac{2\pi\hbar^{2}\tilde{a}_{\text{3D}}n_{0}^{2}}{m}\Big[1\!+\!\frac{\tilde{a}_{\text{3D}}}{\tilde{a}_{\text {cr}}}\!+\!\frac{128}{15}(\frac{n_{0}\tilde{a}_{\text{3D}}^{3}}{\pi})^{1/2}\!\!\frac{m^{*}}{m\left(1+\chi\mu\right)^{2}}\Big]\!.\label{3DGroundEnergy}
\end{eqnarray}
In Equation (\ref{3DGroundEnergy}), the first term of the right-hand side represents the mean-field energy characterized by the lattice-renormalized $s$-wave scattering length of $\tilde{a}_{\text{3D}}=a_{\text{3D}}d^{2}/2\pi\sigma^{2}$. The term of $\tilde{a}_{\text{3D}}/\tilde{a}_{\text {cr}}<0$ is a further renormalization of the scattering length due to the combined effects of the optical lattice and finite-range interaction.  In more details, the $\tilde{a}_{\text {cr}}$ is calculated to be $\tilde{a}_{\text {cr}}=-\frac{d\sqrt{m}}{2\sqrt{2}\left(1.43+9.8\chi\mu\right)\cdot\sqrt{m^{*}}}$. Here, the $m^*=\hbar^{2}/J d^{2}$ is the effective mass associated with the band and the $\chi=4m\tilde{g}_{2}/\hbar^{2}\tilde{g}_{0}$ is due to the finite-range effect. Here, $\tilde{g}_{0}=4\pi\hbar^{2}\tilde{a}_{\text{3D}}/m$ and $\tilde{g}_{2}=2\pi\hbar^{2}\tilde{a}_{\text{3D}}^{2}\tilde{r}_{\text{3D}}/m$ are characterized by the lattice-renormalized $s$-wave scattering length of $\tilde{a}_{\text{3D}}=a_{\text{3D}}d^{2}/2\pi\sigma^{2}$, and the lattice-renormalized effective range of $\tilde{r}_{\text{3D}}=2\pi\sigma^{2}r_{\text{3D}}/d^{2}$.  Note that the $\tilde{a}_{\text{cr}}$ can recover the corresponding one in Ref. \cite{Orso2006} in the vanishing finite-range effect of $\tilde{g}_2=0$.

The last term in Eq. (\ref{3DGroundEnergy}) represents the generalized LHY correction in the presence of the optical lattice and the finite-range interaction. We have $\textcolor{blue}{\text{checked}}$ that  Eq.~(\ref{3DGroundEnergy}) is simplified into the corresponding result in Ref.~\cite{Orso2006} for vanishing the finite-range interaction coupling constant $\tilde{g}_{2}=0$; whereas for vanishing the optical lattice $t=0$, our result recovers exactly the corresponding one in Ref.~\cite{Cappellaro2017}. We see that with respect to the previous cases in Refs. \cite{Orso2006,Cappellaro2017}, the LHY correction in Eq. (\ref{3DGroundEnergy}) is magnified by the renormalization of both the coupling constant of $\tilde{a}_{3D}$, the effective mass of $m^*$ and the finite-range coupling of $\chi=4m\tilde{g}_{2}/\hbar^{2}\tilde{g}_{0}$.

In the opposite 1D regime corresponding $\textcolor{blue}{\text{to}}$ $s\rightarrow 0$, $f\left(s\right)$ exactly approaches a limiting value of $f(0)=4\sqrt{2}/3\left(1+\chi\mu\right)$ with the different values of the finite-range coupling constant of $\chi=4m\tilde{g}_{2}/\hbar^{2}\tilde{g}_{0}$.  We routinely check that our calculated $f(0)=4\sqrt{2}/3$ in the case of vanishing $\chi=0$, which can recover the corresponding one in Ref.~\cite{Orso2006}. In the 1D limit, Equation~(\ref{GroundEnergy2}) approaches asymptotically to the ground state energy of a 1D Bose gas in the presence of the optical lattice and the finite-range effect
\begin{equation}
\frac{E_{g}}{L}=\frac{1}{2}g_{\text{1D}}n_{\text{1D}}^{2}-\frac{2}{3\pi}\textcolor{blue}{\sqrt{\frac{m}{\hbar^2}}}\left(n_{\text{1D}}g_{\text{1D}}\right)^{\frac{3}{2}}\frac{1}{1+\chi\mu},\label{1DGroundEnergy}
\end{equation}
with $n_{\text{1D}}=n_{0}d^{2}$ being the linear density and $L$ being the length of the tube. \textcolor{blue}{It is noteworthy to mention that Eq.~(\ref{1DGroundEnergy}) exhibits a resemblance to the ground state energy of the pure 1D Bose gas in Ref.~\cite{ACappellaro2017}. However, our study incorporates the presence of an optical lattice and derives the asymptotic energy expression in the 1D limit. The zero-range interaction coupling constant $g_{\text{1D}}$ in Eq.~(\ref{1DGroundEnergy}) represents the outcome after renormalization. It is indicated that our findings align well with the pure one-dimensional results in the 1D limit.} In the case of vanishing the finite range interaction coupling constant $\chi=0$, Equation~(\ref{1DGroundEnergy}) is in agreement with the exact LL solution of the 1D model expanded in the weak coupling regime $mg_{\text{1D}}/\hbar n_{\text{1D}}\ll1$~\cite{Lieb1963,Lieb19631616}. We conclude, therefore, that Eq.~(\ref{1DGroundEnergy}) is the generalized LL solution of the 1D model expanded in the weak coupling regime in the presence of the finite range interaction. 

In a similar $\textcolor{blue}{\text{study}}$, we delve into the asymptotic behavior of quantum depletion. In the 1D limit of $s\rightarrow 0$, Equqtion~(\ref{Depletion}) diverges as it is expected, indicating that in the absence of tunneling there is no real Bose-Einstein condensation in alignment with the general theorems in one dimension~\cite{Pitaevskii2016}. In the opposite 3D regime of $s\gg 1$, the function of  $h\left(s\right)$ in Eq. (\ref{ScalingFunH}) decays as $\frac{8}{3\sqrt{2}\pi s}\cdot\left(1-3\chi\mu\right)+\frac{4\cdot\left(\chi\mu\right)^{\frac{3}{2}}}{3 s}$. Consequently, one finds the analytical expression of the quantum depletion in 3D limit
\begin{equation}
n-n_{0}=\frac{8}{3}\sqrt{\frac{\left(\tilde{a}_{\text{3D}}n_{0}\right)^{3}}{\pi}}\frac{m^{*}}{m}\Big[1-3\chi\mu+\frac{\sqrt{2}}{2}\pi\left(\chi\mu\right)^{\frac{3}{2}}\Big].\label{3DDepletion}
\end{equation}
\textcolor{blue}{Replace $\chi$ and $\mu$ in Eq.~(\ref{3DDepletion}) with $4m \tilde{g}_2/ \hbar^2 \tilde{g}_{0}$ and $\tilde{g}_0 n_0$, and rewrite Eq.~(\ref{3DDepletion}) as
	\begin{eqnarray}
		n-n_{0}
		&=&n_{0}\Big[\frac{8}{3}\frac{m^{*}}{m}\sqrt{\frac{\tilde{a}_{\text{3D}}^{3}n_{0}}{\pi}}-64\frac{m^{*}}{m}\sqrt{\pi}\frac{\tilde{r}_{\text{3D}}}{\tilde{a}_{\text{3D}}}\left(\tilde{a}_{\text{3D}}^{3}n_{0}\right)^{\frac{3}{2}}\nonumber\\
		&+&\frac{128}{3}\frac{m^{*}}{m}\pi^{2}\left(\frac{\tilde{r}_{\text{3D}}}{\tilde{a}_{\text{3D}}}\right)^{\frac{3}{2}}\left(\tilde{a}_{\text{3D}}^{3}n_{0}\right)^{2}\Big],\label{3DDepletion2}
\end{eqnarray}}
which generalizes the standard 3D result in the presence of finite range interaction in free space~\cite{Tononi2018} to that in the presence of an optical lattice and finite-range effect. 

\section{CONCLUSION}\label{conclu}

In conclusion, we have meticulously derived the equation of state that extends beyond the mean field approximation for a Bose gas with finite-range interactions within a two-dimensional optical lattice. This derivation was achieved using the framework of finite-temperature functional path integrals. Consequently, we have obtained analytical expressions for ground state energy and quantum depletion. Notably, our work $\textcolor{blue}{\text{has}}$ taken into account finite-range interaction, leading to the dimensional crossover of the system due to constraints on both interaction energy and particle kinetic energy. This crossover differs from the previously observed dimensional crossover in BEC systems, which was solely attributed to the compression of kinetic energy. Our findings provide a more comprehensive understanding of the Bose gas behavior in optical lattices, incorporating the effects of finite-range interactions. Furthermore, in accordance with the Mermin-Wagner-Hohenberg theory~\cite{Mermin1966,Hohenberg1967,Fischer2002}, quantum fluctuations are significantly enhanced by finite-range corrections in low-dimensional systems. It is due to these reasons that our final results are not universal results $\textcolor{blue}{\text{that}}$ solely depend on $\sqrt{n_{0}a_{s}^{3}}$, as previously believed. Our results demonstrated that the two dimensional optical lattice induces the 3D-1D dimensional crossover in quantum fluctuations. In the 1D limit region, our results closely align with the LL solution, offering a precise representation. Conversely, in the 3D limit region, we obtained the generalized LHY correction. These finite-range analytical results represent nontrivial generalizations of the universal equation of state presented in~\cite{Orso2006}. 

The underlying physics of this dimensional crossover involves the intricate interplay among three parameters: the lattice intensity $t$, the effective range $r_{s}$ and the scattering length $a_{s}$. Leveraging the most advanced techniques, all of these quantities are experimentally controllable, allowing for precise manipulation and observation of the system's behavior. Notably, it is feasible to arbitrarily tune the depth of optical lattices ranging from $0E_R$ to $32E_R$~\cite{Du2010}. Therefore, given current experimental capabilities, it should be feasible to directly observe the phenomena discussed herein, which are intimately linked to dimensional effects. The direct observation of this dimensional effect would mark a significant milestone in elucidating the intricate interplay between dimensionality and quantum fluctuations in low dimensions.

We thank Kangkang Li and Ying Hu for stimulating discussions. This work was supported by the National Natural Science Foundation of China (Nos. 12074344), the Zhejiang Provincial Natural Science Foundation (Grant Nos. LZ21A040001) and the key projects of the Natural Science Foundation of China (Grant No. 11835011).

\bibliography{ref}

\end{document}